\documentstyle[aps,preprint,epsfig]{revtex}
\begin{document}
\draft
\tightenlines

\title{Parry measure and the topological entropy of chaotic 
repellers embedded within chaotic attractors}

\author{Hrvoje Buljan\footnote{
Corresponding author: Hrvoje Buljan; e-mail:hbuljan@phy.hr; 
tel: +385-1-460-5650; fax: +385-1-468-0336
} 
and Vladimir Paar}
\address{Department of Physics, Faculty of Science, University of
Zagreb, PP 332, 10000 Zagreb, Croatia}
\date{\today}

\maketitle

\begin{abstract}
We study the topological entropy of chaotic repellers formed 
by those points in a given chaotic attractor that never visit 
some small forbidden hole-region in the phase space. 
The hole is a set of points in the phase space that 
have a sequence $\alpha=(\alpha_0\alpha_1\ldots\alpha_{l-1})$ as 
the first $l$ letters in their itineraries. 
We point out that 
the difference between the topological entropies of 
the attractor and the embedded repeller is for 
most choices of $\alpha$ approximately equal to the 
Parry measure corresponding to $\alpha$, $\mu_P(\alpha)$. 
When the hole encompasses a point of a short periodic orbit, 
the entropy difference is significantly smaller than $\mu_P(\alpha)$. 
This discrepancy is described by the formula which 
relates the length of the short periodic orbit, 
the Parry measure $\mu_P(\alpha)$, and the topological entropies 
of the two chaotic sets. 
\end{abstract}




\section{Introduction}

Besides local instability, the main ingredient of 
chaotic motion is orbit 
complexity\cite{Kat,C_v,O_b}. 
The topological entropy measures orbit complexity 
by looking at all possible motions without regard to 
their likelihood\cite{Kat,C_v,O_b,Adler,Cv,Zyck}. 
Within the field of communicating with chaos, the topological 
entropy of chaotic repellers embedded within chaotic 
attractors is an important issue for the 
analysis of trade-off between the channel capacity and 
noise resistance\cite{BLG,JOH,ZB,LiYi,Boll}.

Motivated by this issue, we 
investigate the following problem. 
We consider an unimodal chaotic map on the interval, 
$f(x):I\rightarrow I$, with the chaotic attractor 
$A\subset I$\cite{Kat,C_v,O_b,Colle,Miln}. 
Imagine a small region $H\subset I$, referred to as the hole, 
located somewhere on the chaotic attractor. 
There is a set of points on $A$ yielding trajectories that 
never enter the hole $H$. 
This set of points is a chaotic repeller (call it $R_H$) 
embedded within the attractor $A$\cite{BLG,JOH,ZB}.

The repeller $R$ governs the dynamics of the 
transiently chaotic map with a hole 
$f_H(x):I\backslash H\rightarrow I$\cite{ZB,LiYi,Chern,B1,PP}. 
The points that map into the hole under the dynamics of the 
map $f_H$ are no longer considered. 
Besides the field of communicating with 
chaos\cite{BLG,JOH,ZB,LiYi}, 
such maps with the hole(s) arise in the study of 
conditionally invariant measures\cite{Chern}, and the 
average lifetimes of chaotic transients preceding 
controlled periodic motion\cite{B1}. 
Here we are interested in the difference between the 
topological entropies of the attractor and the repeller 
for holes $H$ that are small in size.

The critical point $x_c$ of the unimodal map $f$ defines 
the generating partition. 
Every point $x\in I$ is assigned an infinite sequence of 
symbols $\omega(x)\in \Sigma=\{(\omega_0\omega_1\ldots)|\ 
\omega_i\in\{ L,C,R \} \mbox{ for } i\geq 0 \}$ 
by the following prescription: 
if $f^i(x)<x_c$ ($f^i(x)>x_c$), then $\omega_i(x)=L$($R$, respectively); 
if $f^i(x)=x_c$, then $\omega_i(x)=C$\cite{Kat}. 
This definition includes the present state of the system. 
The symbols $\omega_i$ are called letters from the alphabet 
${\mathcal A}=\{L,C,R\}$, and the infinite sequence 
$\omega(x)$ is called the itinerary of the 
point $x$\cite{Kat,C_v,Hansen}. 
Let $\Sigma_A\subset \Sigma$ ($\Sigma_{R_H}\subset \Sigma$) denote 
the set of itineraries corresponding to the trajectories on the 
attractor $A$ (repeller $R_H$, respectively). 
The dynamics on the attractor $A$ (repeller $R_H$) is 
symbolically represented by the topological dynamical 
system $(\Sigma_A,\sigma)$ [$(\Sigma_{R_H},\sigma)$, respectively], 
where $\sigma$ denotes the shift operation\cite{Kat,C_v,Hansen}

\begin{equation}
\sigma(\omega_0\omega_1\omega_2\ldots)=
(\omega_1\omega_2\omega_3\ldots).
\end{equation}
The topological entropy of the attractor $A$ (repeller $R$), 
denoted by $h_A$ ($h_R$, respectively), is identical to the 
topological entropy of $(\Sigma_A,\sigma)$ 
[$(\Sigma_{R_H},\sigma)$, respectively]. 
Since the sets $A$ and $R_H$ are invariant under $f$, 
$\Sigma_A$ and $\Sigma_{R_H}$ are shift-invariant.

We will simplify the analysis to allow for analytical calculations 
by special tailoring of the hole $H$. 
Let $H\equiv H_{\alpha}$ be the set of all points $x\in I$ that have the 
sequence $\alpha=(\alpha_0\alpha_1\ldots\alpha_{l-1})$ 
as the first $l$ letters in their itineraries
$\alpha_i\in \{L,R\}$ 
(see Ref. \cite{Arrow} for equivalent tailoring 
in the context of controlling chaos). 
The endpoints of the interval $H_{\alpha}$ are 
preimages of the critical point. 
The itinerary of some point $x$ contains the sequence 
$(\alpha_0\alpha_1\ldots\alpha_{l-1})$ if and only if 
$x$ eventually visits the hole $H_{\alpha}$. 
Hence, 
the set $\Sigma_{R_H}$ is obtained from the set $\Sigma_A$ 
by extracting from it all itineraries that contain the sequence 
$\alpha$.

Among all measures with support in $\Sigma_A$ that 
are shift-invariant, Parry measure of the 
dynamical system $(\Sigma_A,\sigma)$ maximize the value 
of entropy\cite{Kat,Parry}; the entropy of $(\Sigma_A,\sigma)$ 
with respect to the Parry measure $\mu_P$ is 
equal to the topological entropy\cite{Kat}. 
If the Parry measure contained within the cylinder 
$C^l_{\alpha}=\{\omega\in \Sigma |\ 
\omega_i=\alpha_i \mbox{ for } i=0,1,\ldots,l-1 \}$ is larger, 
the sequence $\alpha$ more frequently appears within 
itineraries $\omega\in\Sigma_A$. 
Hence, it is reasonable to expect that the entropy difference 
$\Delta_h(\alpha):=h_A-h_R(\alpha)$ will be larger 
for larger values of $\mu_P(C^l_{\alpha})$. 
[From now on, we abbreviate $\mu_P(\alpha):=\mu_P(C^l_{\alpha})$.] 
In this manuscript we study the correlation between the 
Parry measure $\mu_P(\alpha)$, and the entropy difference 
$\Delta_h(\alpha)$, for various choices of $H_{\alpha}$.

We will numerically demonstrate that 
if the hole $H_{\alpha}$ is not visited by a short periodic 
orbit, then $\Delta_h(\alpha)$ is approximately equal to the 
Parry measure $\mu_P(\alpha)$. 
By using this observation, it will be demonstrated, 
analytically and numerically,
that when $H_{\alpha}$ is visited by a short periodic orbit, 
then $\Delta_h(\alpha)$ is significantly smaller than $\mu_P(\alpha)$. 
The ratio, $\mu_P(\alpha)/\Delta_h(\alpha)$, which 
characterizes this discrepancy, can be expressed in terms of the 
topological entropy $h_A$, and the length (call it $p$) 
of the shortest periodic orbit visiting $H_{\alpha}$: 

\begin{equation}
\frac{\mu_P(\alpha)}
{\Delta_h(\alpha)}\simeq \frac{1}{1-e^{-ph_A}}. 
\label{final}
\end{equation}
Formula (\ref{final}) is derived by using the following auxiliary result. 
Suppose that $x$ is a periodic orbit of prime period $p$, 
and $\alpha_i=\omega_i(x)$, for $i=0,1,\ldots,l-1$. 
For $l\gg 1$, the entropy difference $\Delta_h(\alpha)$, 
and $\mu_P(\alpha)$ decrease exponentially fast with the 
increase of $l$. 
The characteristic exponent is shown to be approximately 
equal to the topological entropy $h_A$.

\section{Numerical and theoretical analysis of the 
entropy difference and the Parry measure}

For the illustrations of the dependence of 
$\Delta_h(\alpha)$ against the position of $H_{\alpha}$ on 
the attractor, it is unconvenient to draw $\Delta_h(\alpha)$ 
against the mid-point of the hole $H_{\alpha}$. 
Namely, the holes $H_{\alpha}$ vary in size for different 
unimodal maps. 
Furthermore, the intervals $I$ may differ for different maps. 
For these reasons, it is convenient to introduce the 
topological mid-point of the hole $H_{\alpha}$. 
To a sequence $\alpha=(\alpha_0\ldots\alpha_{l-1})$, we assign a 
number

\begin{equation}
t(\alpha)=\sum_{i=0}^{l-1}
\frac{\gamma_i}{2^{i+1}}+\frac{1}{2^{l+1}},
\label{spao}
\end{equation}
where $\gamma_i=0(1)$ if the number of $R$s in the 
sequence $(\alpha_0\alpha_1\ldots\alpha_i)$ is even 
(odd, respectively). 
Given a sequence $\alpha$, $t(\alpha)$ is identical 
for all unimodal maps, and it is called the topological mid-point 
of $H_{\alpha}$ since it preserves the spatial ordering of holes: 
if $t(\alpha')<t(\alpha'')$, then $H_{\alpha'}$ is to the 
left of $H_{\alpha''}$\cite{Spa}.

In order to illustrate the main result of this paper, 
consider the attractor $A=[0,1]$ of the tent map 
$T_{a}(x)=a(1-|1-2x|)/2$ for $a=2$. 
The topological entropy is $h_A=\ln 2$. 
Every sequence is admissible by the systems dynamics, i.e., 
$(\Sigma_A,\sigma)$ is a shift of full type on two 
symbols\cite{Bin}. 
The Parry measure corresponding to any sequence of 
length $l$ is $1/2^l$, i.e., $\mu_P(\alpha)=1/2^l$. 
Figure \ref{fig2}(a) displays $\Delta_h(\alpha)$ and 
$\mu_P(\alpha)$, while Fig. \ref{fig2}(b) displays the ratio 
$\mu_P(\alpha)/\Delta_h(\alpha)$ against $t(\alpha)$. 
We observe that for almost every choice of $\alpha$, 
$\Delta_h(\alpha)\simeq \mu_P(\alpha)$. 
However, when the hole $H_\alpha$ is visited by a short periodic 
orbit of period $p$ smaller than some critical value 
$l_c\sim 3-4$, the quantity $\mu_P(\alpha)$ is significantly 
larger than $\Delta_h(\alpha)$.

Consider the map $T_{a}(x)$ for $a=1.72208\ldots$, with 
the topological entropy $h_A=\ln a=0.543535\ldots$. 
The dynamics is pruned, that is, there are sequences 
inadmissible by the dynamics of the map $T_{1.72208\ldots}(x)$, and 
$(\Sigma_A,\sigma)$ is a shift of finite type on two symbols. 
Fig. \ref{fig3} illustrates $\mu_P(\alpha)$, $\Delta_h(\alpha)$, 
and their ratio against $t(\alpha)$ for 
all admissible sequences of length $l$. 
The same observations as in Fig. \ref{fig2} can be made.

After performing a number of numerical experiments equivalent to 
the ones illustrated in Figs. \ref{fig2} and \ref{fig3}, 
we make the following conjecture: 
If $H_{\alpha}$ is not visited by a short periodic orbit, 
then $\mu_P(\alpha)\simeq \Delta_h(\alpha)$. 
By using this conjecture, we can analytically describe 
the discrepancy between $\mu_P(\alpha)$ and $\Delta_h(\alpha)$ 
which occurs when $H_{\alpha}$ encompasses a short periodic orbit.

At first, we must study the dependence of $\Delta_h(\alpha)$ 
and $\mu_P(\alpha)$ on the length of the sequence $\alpha$. 
We perform the following numerical experiment. 
Consider a point $x\in A$, which lies on a periodic orbit 
of prime period $p$. 
The itinerary $\omega(x)=\pi^{\infty}$ is defined by the 
sequence of $p$ letters, $\pi=(\pi_0\pi_1\ldots\pi_{p-1})$.  
By choosing

\begin{equation}
\alpha=(\underbrace{\pi_0\pi_1\ldots\pi_{p-1}}_{0}
\underbrace{\pi_0\pi_1\ldots\pi_{p-1}}_{1}
\ldots
\underbrace{\pi_0\pi_1\ldots\pi_{p-1}}_{r-1}
\pi_0\pi_1\ldots\pi_{r'-1}
),
\label{alphcon}
\end{equation}
where $r=l \mbox{ div } p$, and $r'=l \mbox{ mod } p$, 
the hole $H_{\alpha}$ encompasses the point $x$ for 
every $l=rp+r'$. 
(The size of the hole $H_{\alpha}$ decreases with the increase of $l$.) 
Fig. \ref{fig4} illustrates the dependence of the difference 
$\Delta_h(\alpha)$, and the ratio $\mu_P(\alpha)/\Delta_h(\alpha)$ 
on the length $l$ of the sequence $\alpha$; 
$\alpha$ is constructed from $\pi=(R)$, $(RL)$, and 
$(LLR)$ [see Eq. (\ref{alphcon})]. 
For this illustration, we have utilized the map $T_{a}(x)$ 
for $a=2$, and $a=1.72208\ldots$.

We observe the following.
The difference $\Delta_h(\alpha)$ decreases 
approximately exponentially fast with the increase of $l$. 
The points in Figs. \ref{fig4}(a), and \ref{fig4}(b) 
are fitted to the functional dependence 
$\Delta_h(\alpha)=A_0e^{-A_1l}$. 
The fitted values of the exponent $A_1$ are shown in Table \ref{tab1}. 
The characteristic exponent $A_1$ is 
approximately equal to the topological entropy $h_A$. 
In some cases, the graph $\Delta_h(\alpha)$ against $l$ 
shows flat plateaus, as is illustrated in Fig. 
\ref{fig4}(b). 
From Fig. \ref{fig4}(c) we observe that 
the ratio $\mu_P(\alpha)/\Delta_h(\alpha)$ is independent of $l$ 
for $l$ large enough. 
In other words, both $\Delta_h(\alpha)$ and $\mu_P(\alpha)$ are 
observed to have the same (up to the multiplicative constant) 
functional dependence on $l$, 
$\Delta_h(\alpha)\sim \mu_P(\alpha) \sim e^{-h_Al}$.

In order to explain these observations, we utilize the 
topological zeta function\cite{C_v,Artin}. 
Let $\alpha$ and $\beta$ denote two sequences 
of length $l$ and $L\gg l$, respectively, which 
are constructed as prescribed in Eq. (\ref{alphcon}), 
so that $\beta_i=\alpha_i$ for $i=0,1,\ldots,l-1$. 
These sequences define the embedded repellers 
$R^{(\alpha)}$ and $R^{(\beta)}$. 
Let $q$ denote the length of the shortest periodic 
orbit(s) that visits the hole $H_{\alpha}$, and 
which does not visit the hole $H_{\beta}\subset H_{\alpha}$. 
This orbit is a part of the repeller $R^{(\beta)}$, but it does not 
belong to $R^{(\alpha)}$. 
The topological entropies of $R^{(\alpha)}$ and $R^{(\beta)}$ 
can be calculated from the smallest positive zeros of their 
corresponding topological zeta functions\cite{C_v}: 

\begin{equation}
1/\zeta_{top}^{(\alpha)}=1-\sum_{n=1}^{q-1}c_n z^n
-\sum_{n=q}^{\infty}c_n^{(\alpha)} z^n,
\label{zeta1}
\end{equation}
and

\begin{equation}
1/\zeta_{top}^{(\beta)}=1-\sum_{n=1}^{q-1}c_n z^n
-\sum_{n=q}^{\infty}c_n^{(\beta)}z^n. 
\label{zeta2}
\end{equation}
Only the coefficients of order $q$ and higher differ 
since the periodic orbits within the repellers $R^{(\alpha)}$ 
and $R^{(\beta)}$ that are of prime period $q-1$ and smaller 
coincide\cite{C_v}. 
For $H_{\alpha}$, and $H_{\beta}$ as constructed above, 
it can be shown that $q>(r-1)p$, and most likely 
$q\leq (r+1)p$, that is, $q\sim l=rp+r'$ (see Appendix A).

Let us define the function 
$\Delta_{\zeta}(z):=1/\zeta_{top}^{(\beta)}-1/\zeta_{top}^{(\alpha)}$, 
and the quantity $\epsilon_z:=z^{(\alpha)}-z^{(\beta)}$, where 
$z^{(\alpha)}$ and $z^{(\beta)}$ denote the smallest positive 
zeros of the functions $1/\zeta_{top}^{(\alpha)}$ and 
$1/\zeta_{top}^{(\beta)}$, respectively. 
The topological entropies of the repellers $R^{(\alpha)}$, 
and $R^{(\beta)}$ are $h^{(\alpha)}=-\ln z^{(\alpha)}$, 
and $h^{(\beta)}=-\ln z^{(\beta)}$, respectively. 
If $l$ is large enough, then $\eta=\epsilon_z/z^{(\beta)}\ll 1$. 
This approximation leads to 

\begin{equation}
h^{(\beta)}-h^{(\alpha)}=
\ln\frac{z^{(\alpha)}}{z^{(\beta)}}= 
\frac{\epsilon_z}{z^{(\beta)}}+{\mathcal O}(\eta^2), 
\label{treca}
\end{equation}
and

\begin{eqnarray}
\Delta_{\zeta}(z^{(\beta)}) & = & 
0-1/\zeta_{top}^{(\alpha)}(z^{(\beta)}) \nonumber \\
 & = & -1/\zeta_{top}^{(\alpha)}(z^{(\alpha)}-\epsilon_z) \nonumber \\
 & = & 
z^{(\beta)}\frac{d[1/\zeta_{top}^{(\alpha)}(z)]}{dz}
\mid_{z=z^{(\alpha)}}\frac{\epsilon_z}{z^{(\beta)}}+{\mathcal O}(\eta^2).
\label{cetvrta}
\end{eqnarray}
On the other hand, from Eqs. (\ref{zeta1}) and (\ref{zeta2})

\begin{eqnarray}
\Delta_{\zeta}(z^{(\beta)}) & = & 
\sum_{n=q}^{\infty}[c_n^{(\alpha)}-c_n^{(\beta)}]
[z^{(\beta)}]^n \nonumber \\
 & = & [z^{(\beta)}]^{q}
\sum_{n=0}^{\infty}[c_{n+q}^{(\alpha)}-c_{n+q}^{(\beta)}]
[z^{(\beta)}]^{n}. 
\label{peta}
\end{eqnarray}
In the limit $L\rightarrow \infty$, $h^{(\beta)}=h_A$, i.e.,
$z^{(\beta)}=e^{-h_A}$. 
By neglecting the terms of order $\eta^2$, and higher, 
from Eqs. (\ref{treca}), 
(\ref{cetvrta}), and (\ref{peta}) we obtain 

\begin{eqnarray}
\ln \Delta_h(\alpha) & \simeq & -qh_A +
\ln \left| 
\sum_{n=0}^{\infty}[c_{n+q}^{(\alpha)}-c_{n+q}^{(\beta)}]
[z^{(\beta)}]^{n}
\right| \nonumber \\
 & - & 
\ln \left| 
z^{(\beta)}\frac{d[1/\zeta_{top}^{(\alpha)}(z)]}{dz}
\mid_{z=z^{(\alpha)}}
\right|. 
\label{just}
\end{eqnarray}
If the sequence $\alpha$ is long enough, 
the first term on the right hand side (r.h.s.) of 
Eq. (\ref{just}), $-qh_A\sim -lh_A$, is much larger 
than the second, and the third term. 
This explains the observed functional 
dependence $\Delta_h(\alpha)\sim e^{-lh_A}$. 
The second term on the r.h.s. can introduce 
small oscillations superimposed on the global behavior 
$\Delta_h(\alpha)\sim e^{-lh_A}$ [e.g., see open circles in 
Fig. \ref{fig4}(a)]. 
The third term is practically independent of $l$ for 
large enough $l$. 
Eq. (\ref{just}) is consistent with the 
scaling of $h_A$ with the size of the hole that 
is reported in Ref. \cite{JOH} for the tent, 
$2x \mbox{ mod } 1$, and the logistic map $4x(1-x)$.

The dependence of  $\Delta_h(\alpha)$ on 
$l$ is equivalent to the problem of calculating 
the topological entropy of some attractor with infinite Markov 
partition by using the set of finite pruning rules. 
By using the set of finite pruning rules, one 
actually calculates the topological entropy of the 
embedded repeller\cite{C_v}. 
If all periodic orbits up to the length $l$ are 
included in the calculation of the topological entropy of the 
attractor, then the difference between the topological entropies 
of the embedded repeller and the attractor is of order 
$\sim e^{-lh_A}$ (see the discussion in Ref. \cite{C_v}).

In Refs. \cite{BLG,JOH,ZB,LiYi} the entropy $h_R$ has been 
studied as a function of the size of the hole $H$. 
It has been demonstrated that $h_R$ is a devil's 
staircase like function with flat plateaus. 
In the case considered here, the size of the hole 
is not a continuous parameter since it 
enters the calculations via the length 
$l$ of the sequence $\alpha$. 
The flat plateaus in the graph $\Delta_h(\alpha)$ vs. 
$l$ [see Fig. \ref{fig4}(b)] are a fingerprint of the 
plateaus observed and explained in 
Refs. \cite{BLG,JOH,ZB,LiYi} in a more general settings.

If $(\Sigma_A,\sigma)$ is $n$-step topological Markov chain, 
the explanation of the functional dependence 
$\mu_P(\alpha)\sim e^{-lh_A}$ follows immediately 
from the definition of the Parry measure 
(see Appendix B, and Ref. \cite{Kat}, p. 175). 
For a more general class of topological dynamical systems 
$(\Sigma_A,\sigma)$, which naturally arise from 
one dimensional maps, the functional dependence 
$\mu_P(\alpha)\sim e^{-lh_A}$ is explained by using 
the following property of the map $T_a(x)$. 
Consider an unimodal map $f$ with $h_A>0$, and the 
corresponding topological dynamical system $(\Sigma_A,\sigma)$. 
The map $T_a(x)$ with $a=e^{h_A}$ corresponds to the same 
topological dynamical system $(\Sigma_A,\sigma)$ 
(see Ref. \cite{Kat}, p. 514.). 
Let $h_{\mu_N}$ denote the entropy of $T_a(x)$ with 
respect to the naturally invariant measure $\mu_N$. 
The entropy $h_{\mu_N}$ is equal to the Lyapunov 
exponent $\lambda=\ln a$\cite{O_b,BR}, i.e., 
for the map $T_{a}(x)$ we have 
$h_{\mu_N}\equiv \lambda \equiv h_A=\ln a$. 
Since $h_A$ is the entropy with respect to the 
maximal entropy measure\cite{Kat}, we conclude that 
the maximal entropy measure (i.e. the Parry measure) 
coincides with the naturally invariant measure $\mu_N$. 
Therefore, $\mu_P(\alpha)\equiv \mu_N(H_\alpha)$. 
The length of the interval $H_{\alpha}$ is roughly 
$\sim 1/a^l$. 
From the scaling of $\mu_N$ it follows that\cite{O_b}, 

\begin{equation}
\mu_P(\alpha)\equiv \mu_N(H_\alpha)\sim 1/a^l=e^{-lh_A}.
\end{equation}
Equivalent analysis demonstrates that Parry measure 
coincides with the naturally invariant measure 
on the chaotic repeller of the properly chosen 
gap-tent map (see. Sec. 2 in Ref. \cite{ZB}).

By using two points from the previous discussion: 
(i) The fact that  
$\mu_P(\alpha)\sim e^{-lh_A}$, and 
(ii) the observation that for $H_{\alpha}$ 
which does not encompass a short periodic orbit 
$\Delta_h(\alpha)\simeq \mu_P(\alpha)$, 
we can explain the discrepancy between $\mu_P(\alpha)$ 
and $\Delta_h(\alpha)$ which occurs when $H_{\alpha}$ is visited 
by a short periodic orbit. 
We will derive Eq. (\ref{final}) for $H_{\alpha}$ 
located on a fixed point, i.e., for $\alpha=(R^l)$, 
$p=1$. 
The generalization for the case when $H_{\alpha}$ 
encompasses a point on a short periodic orbit of prime period 
$p=2,3,\ldots,l_c$, is straightforward.

The first preimages of the cylinder $C^{l}_{(R^l)}$ are 
the cylinders $C^{l+1}_{(LR^l)}$ and 
$C^{l+1}_{(RR^l)}=C^{l+1}_{(R^{l+1})}$. 
From the invariance of Parry measure it follows that

\begin{equation}
\mu_P(R^l)=\mu_P(LR^l)+\mu_P(R^{l+1}). 
\end{equation}
Since $\mu_P(R^{l+1})/\mu_P(R^l)=e^{-h_A}$, 

\begin{equation}
\mu_P(R^l)\simeq (1-e^{-h_A})^{-1}\mu_P(LR^l). 
\label{prva}
\end{equation}
Let us compare $\Delta_h(R^l)$, and $\Delta_h(LR^l)$. 
Let $P_{(R^l)}$ [$P_{(LR^l)}$] denote the set of all 
periodic orbits within the attractor $A$ which contain the sequence 
$(R^l)$ [$(LR^l)$, respectively]. 
It is easy to see that every orbit within $P_{(R^l)}$, except 
the fixed point $(R^{\infty})$ itself, contains the sequence 
$(LR^l)$. 
Thus, the two sets differ by a single periodic orbit. 
Since the topological entropy quantitatively expresses 
the exponential growth of periodic orbits, 
it follows that $h_R(R^l)$ and $h_R(LR^l)$ are identical, 
i.e., 

\begin{equation}
\Delta_h(R^l)=\Delta_h(LR^l).
\label{druga}
\end{equation}
By combining Eqs. (\ref{prva}) and (\ref{druga}), and using the 
observation $\mu_P(LR^l)\simeq \Delta_h(LR^l)$, we 
immediately obtain Eq. (\ref{final}) for the case $p=1$. 
To generalize for the case $p=2,3,\ldots$, one has to 
consider the $p$th order preimages of the cylinder 
$C^{l}_{\alpha}$ to obtain equations equivalent to 
(\ref{prva}) and (\ref{druga}). 
Fig. \ref{fig4}(c) confirms the validity of the approximations 
that lead to Eq. (\ref{final}).
These approximations work 
better for larger values of $l$ [see Fig. \ref{fig4}(c)].

\section{Conclusion}

We have studied the topological entropy of an 
embedded chaotic repeller $R_H$, defined as the set 
of points on a given attractor $A$ that do not 
visit a small region (hole) in the phase space $H_{\alpha}$. 
The hole $H_{\alpha}$ is a set of points that 
have a sequence $\alpha=(\alpha_0\alpha_1\ldots\alpha_{l-1})$ as 
the first $l$ letters in their itineraries. 
It has been shown that the difference between the 
topological entropies of the attractor and the repeller, 
denoted by $\Delta_h(\alpha)$, is for most choices of $\alpha$ 
approximately equal to the Parry measure corresponding to $\alpha$, 
i.e., $\Delta_h(\alpha)\simeq \mu_P(\alpha)$. 
Significant deviation from this global behavior occurs when 
the itinerary of some short periodic orbit contains 
the sequence $\alpha$. 
The formula which relates the length of the short periodic orbit, 
the Parry measure $\mu_P(\alpha)$, and the entropy 
difference $\Delta_h(\alpha)$ has been derived [see Eq. (\ref{final})].

\section{Appendix A}

Consider a periodic orbit 
$\omega^{(\pi)}=\pi^{\infty}\in \Sigma_A$ of prime 
period $p$; it is defined by the sequence of $p$ letters, 
$\pi=(\pi_0\pi_1\ldots\pi_{p-1})$. 
Let $\alpha$ and $\beta$ denote two sequences of length $l$ 
and $L\gg l$, respectively, that are constructed from $\pi$ 
[see Eq. (\ref{alphcon})]. 
We assume that $r=(l \mbox{ mod } p)\geq 3$. 
Clearly, $\omega^{(\pi)}\in C^{l}_{\alpha}$. 
Let $\omega^{(\upsilon)}=\upsilon^{\infty}\in \Sigma_A$ denote 
a periodic orbit of prime period $q$, 
defined by the sequence $\upsilon=(\upsilon_0\upsilon_1\ldots\upsilon_{q-1})$,
such that $\omega^{(\upsilon)}\neq \omega^{(\pi)}$, and 
$\omega^{(\upsilon)}\in C^{l}_{\alpha}$; 
$\omega^{(\upsilon)}\in C^{l}_{\alpha}$ means that 
$\omega^{(\pi)}_i=\omega^{(\upsilon)}_i$, for $i=0,1,\ldots,l$. 
In this appendix, we demonstrate that $q>(r-1)p$.

(1) Suppose that $p\leq q \leq (r-1)p$. 
We define integers $m=q \mbox{ div } p$, and $n=q \mbox{ mod } p$ 
($q=mp+n$). 
For $n=0$, $\upsilon=\pi^m$; this implies 
$\omega^{(\upsilon)}=\omega^{(\pi)}$, and we are in contradiction. 
For $n>0$, the comparison of $\alpha$ constructed form $\pi$ 
and $\upsilon$ yields 
$\pi_{(n+i) \mbox{ mod } p}=\alpha_{q+i}=\upsilon_i=\pi_i$ for 
$i=0,1,\ldots p-1$, and this contradicts the assumption 
that $\omega^{(\pi)}$ is of {\em prime} period $p$. 

(2) Suppose that $q<p$, and define integers 
$m=p \mbox{ div } q$, and $n=p \mbox{ mod } q$, $n>0$ ($p=mq+n$). 
Since $\omega^{(\upsilon)}\in C^{l}_{\alpha}$ implies 
$\upsilon_{(n+i)\mbox{ mod } q}=\alpha_{p+i}=\pi_i=\upsilon_{i}$ for 
$i=0,1,\ldots,q-1$, $\omega^{(\upsilon)}$ can not be of prime 
period $q$. 

Thus, we have demonstrated that $q>(r-1)p$. 
Since periodic orbits are dense in $\Sigma_A$\cite{Kat}, 
for a given sequence $\alpha$ constructed from $\pi$, 
there most likely exists a periodic orbit 
$\omega^{(\upsilon)}$ of prime period $q$, 
such that $(r-1)p<q \leq (r+1)p$, and 
$\omega^{(\upsilon)}\in C^l_{\alpha}$. 
For large enough $L\ll 1$, 
this orbit is not within the cylinder $C^l_{\beta}$. 
Therefore, the length $q$ of the shortest periodic orbit that 
visits the hole $H_{\alpha}$, and which does not visit the hole 
$H_{\beta}$ is $q>(r-1)p$, and most likely $q\leq (r+1)p$. 

\section{Appendix B}

In this Appendix, we recall the definition of the 
Parry measure for $n$-step topological Markov chain\cite{Kat,Parry}. 
From this definition the explanation of the functional 
dependence $\mu_P(\alpha)\sim e^{-lh_A}$ follows immediately.

The $2^n\times 2^n$ matrix $C$ which determines all admissible 
transitions is defined as follows\cite{Kat}. 
The matrix element 
$C_{(i_0i_1\ldots i_{n-1}),(j_0j_1\ldots j_{n-1})}$, 
where $i_k,j_k\in \{L,R\}$, is equal to $1$ if 
$j_k=i_{k+1}$ for $k=0,1,\ldots,n-2$, and if 
$(i_1\ldots i_{n-1}j_{n-1})$ is an admissible sequence. 
Otherwise, $C_{(i_0i_1\ldots i_{n-1}),(j_0j_1\ldots j_{n-1})}=0$. 
[The indices of the matrix elements are not integers, 
but sequences of $L$s and $R$s. 
By identifying $L$ ($R$) with $0$ ($1$, respectively), 
the index sequence $(i_0i_1\ldots i_{n-1})$ can be thought of 
as an integer $\in [0,2^n-1]$ represented in the 
binary system].

The largest eigenvalue of the matrix $C$ is identical 
to $e^{h_A}$\cite{Kat,C_v}. 
Let $y$, and $w$ denote the eigenvectors of the matrix 
$C$ and its transpose $C^T$, respectively, 
corresponding to the eigenvalue $e^{h_A}$; these 
eigenvectors are normalized so that 

\begin{equation}
\sum_{(i_0i_1\ldots i_{n-1})}
w_{(i_0i_1\ldots i_{n-1})}
y_{(i_0i_1\ldots i_{n-1})}=1,\ i_k\in \{L,R\}.
\label{norm}
\end{equation}
To find the Parry measure, another 
$2^n\times 2^n$ matrix (call it $P$) is constructed 
(see Ref. \cite{Kat}. p. 175.), 

\begin{eqnarray}
P_{(i_0i_1\ldots i_{n-1}),(j_0j_1\ldots j_{n-1})} & = & 
e^{-h_A}C_{(i_0i_1\ldots i_{n-1}),(j_0j_1\ldots j_{n-1})}\nonumber \\
 & & \frac{w_{(i_0i_1\ldots i_{n-1})}}{w_{(j_0j_1\ldots j_{n-1})}}.
\label{Pdef}
\end{eqnarray}
The Parry measure corresponding to the sequence 
$(i_0i_1\ldots i_{n-1})$ is\cite{Kat} 

\begin{equation}
\mu_P(i_0i_1\ldots i_{n-1})=
w_{(i_0i_1\ldots i_{n-1})}
y_{(i_0i_1\ldots i_{n-1})}. 
\label{pdef}
\end{equation}
The Parry measure corresponding to a sequence 
$\alpha=(\alpha_0\alpha_1\ldots\alpha_{l-1})$, of length 
$l\geq n$ is\cite{Kat}

\begin{eqnarray}
\mu_P(\alpha) & = & P_{(\alpha_0\alpha_1\ldots\alpha_{n-1}),
               (\alpha_1\alpha_2\ldots\alpha_n)}
                    P_{(\alpha_1\alpha_2\ldots\alpha_n),
               (\alpha_2\alpha_3\ldots\alpha_{n+1})}\nonumber \\
 & & \cdots P_{(\alpha_{l-n-1}\alpha_{l-n}\ldots\alpha_{l-2}),
               (\alpha_{l-n}\alpha_{l-n+1}\ldots\alpha_{l-1})}\nonumber \\
 & &  \mu_P(\alpha_{l-n}\alpha_{l-n+1}\ldots\alpha_{l-1}) \nonumber \\
 & = & e^{-(l-n)h_A} 
w_{(\alpha_0\alpha_1\ldots\alpha_{n-1})}
y_{(\alpha_{l-n}\alpha_{l-n+1}\ldots\alpha_{l-1})}
\label{mu_Padef}.
\end{eqnarray}
The last equality follows from the assumption $\mu_P(\alpha)\neq 0$. 
If $\alpha$ is defined as in Eq. (\ref{alphcon}), from 
Eq. (\ref{mu_Padef}) it immediately follows that 
$\mu_P(\alpha)\sim e^{-lh_A}$ with possible oscillations 
of period $p$.

\begin{figure}
\caption{(a) The difference $\Delta_h(\alpha)$ (solid line) and the 
Parry measure $\mu_P(\alpha)$ (dashed line) against $t(\alpha)$. 
(b) The ratio $\mu_P(\alpha)/\Delta_h(\alpha)$ against $t(\alpha)$. 
The integers above the peaks denote the length of 
the shortest periodic orbit visiting $H_{\alpha}$. 
The length of the sequence $\alpha$ is $l=8$. 
}
\label{fig2} 
\end{figure}

\begin{figure}
\caption{(a) The Parry measure $\mu_P(\alpha)$ corresponding to the 
allowed sequences of length $l=8$. 
(b) The difference $\Delta_h(\alpha)$, and  
(c) the ratio $\mu_P(\alpha)/\Delta_h(\alpha)$ 
against the topological mid-point of $H_{\alpha}$. 
The integers above the peaks denote the length of 
the shortest periodic orbit visiting $H_{\alpha}$. 
}
\label{fig3} 
\end{figure}

\begin{figure}
\caption{
(a) The difference $\Delta_h(\alpha)$ against 
the length $l$ of the sequence $\alpha$. 
The lines represent the least square fit to the 
functional dependence $\Delta_h(\alpha)=A_0e^{-A_1l}$. 
(b) The same as figure (a). 
(c) The ratio $\Delta_h(\alpha)/\mu_P(\alpha)$ against $l$. 
The lines represent the quantity $(1-e^{-ph_A})^{-1}$ 
[see Eq. (\ref{final})]. 
Closed squares correspond to the map $T_a(x)$ 
for $a=2$, and $\alpha=(RR\ldots R)=(R^l)$; 
closed circles correspond to $a=2$, and 
$\alpha=(RLRL\ldots)$; 
open squares correspond to $a=1.72208\ldots$, and 
$\alpha=(R^l)$; 
open circles correspond to $a=1.72208\ldots$, and 
$\alpha=(RLRL\ldots)$;
open diamonds correspond to $a=1.72208\ldots$, and 
$\alpha=(LLRLLR\ldots)$. }
\label{fig4} 
\end{figure}

\begin{table}
\begin{tabular}{cccc} 
   a   &   $\pi$  &  $A_1$   & $h_A$   \\
\hline
$2.0$      &  $(R)$ & $0.6958$ &   $0.6931\ldots$      \\
  &  $(RL)$ & $0.6968$ &    \\
\hline
 & $(R)$  & $0.550$ &     \\
$1.72208\ldots$ &  $(RL)$  & $0.565$ &  $0.5435\ldots$  \\
 &  $(LLR)$ & $0.533$ &     
\end{tabular}
\caption{
The fitted values of the parameter $A_1$ 
[$\Delta_h(\alpha)=A_0 e^{-lA_1}$] for the map $T_{a}(x)$, and 
sequences $\alpha$ constructed from $\pi$ 
[see Eq. (\ref{alphcon})]}
\label{tab1}
\end{table}


\begin{references}
\bibitem{Spa} Formula (\ref{spao}) is an adaptation of 
an equivalent formula [see e.g., Eqs. (2.3) and (2.4) in Ref. \cite{Cv}, 
and also Refs. \cite{C_v,Hansen}] that define the ordering of 
itineraries, which preserves the ordering of corresponding 
points for unimodal maps.
\bibitem{Bin} Although the alphabet is ${\mathcal A}=\{L,C,R\}$, 
there is only one itinerary within $\Sigma_A$ that begins with the 
letter $C$, namely the kneading sequence \cite{Kat}. 
Therefore, in the case of unimodal one-dimensional maps, we are 
effectively dealing with the binary alphabet 
(eg., see Ref. \cite{C_v}).
\end{references}
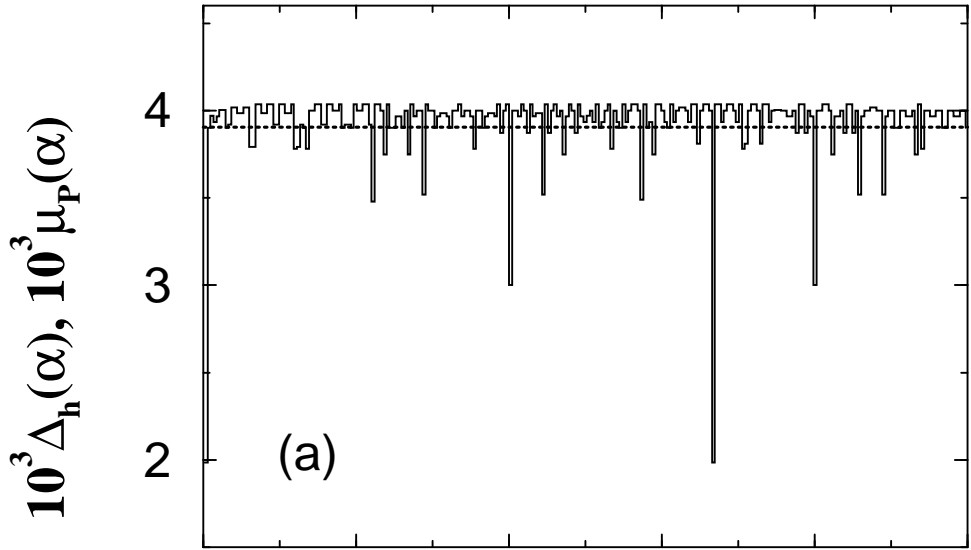
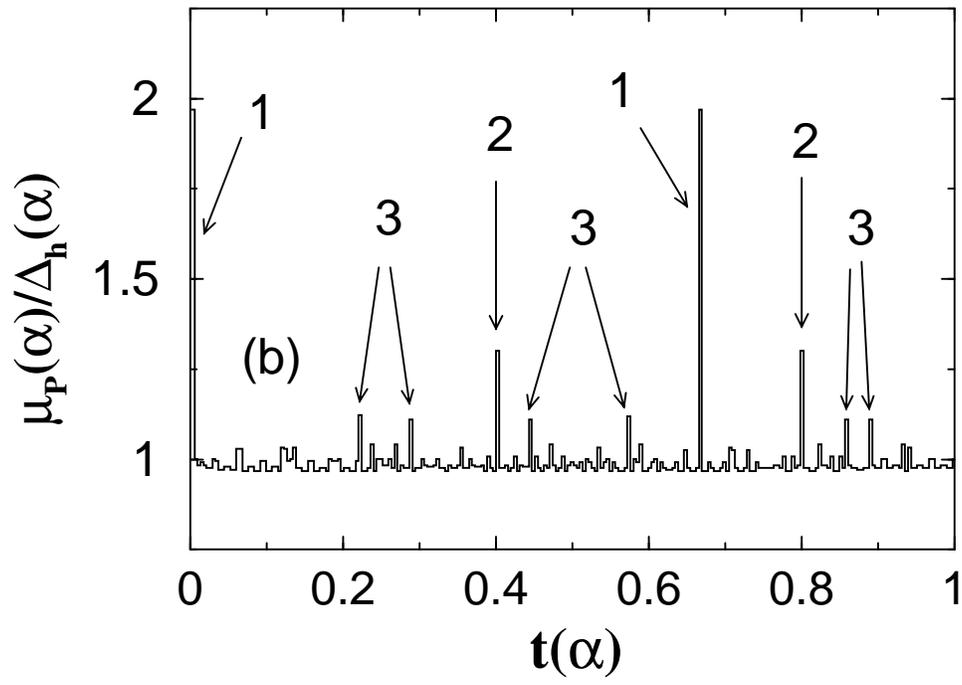
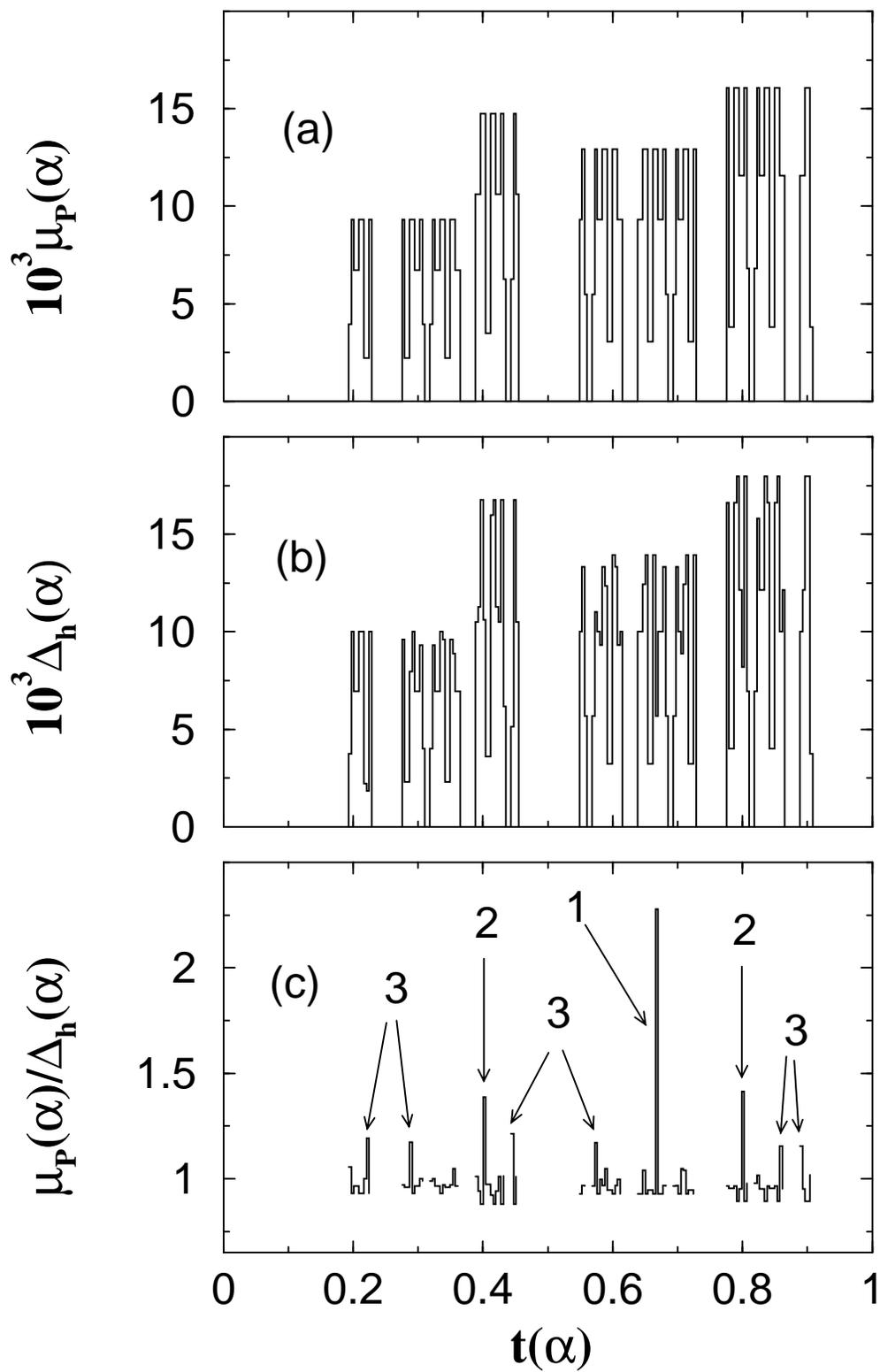
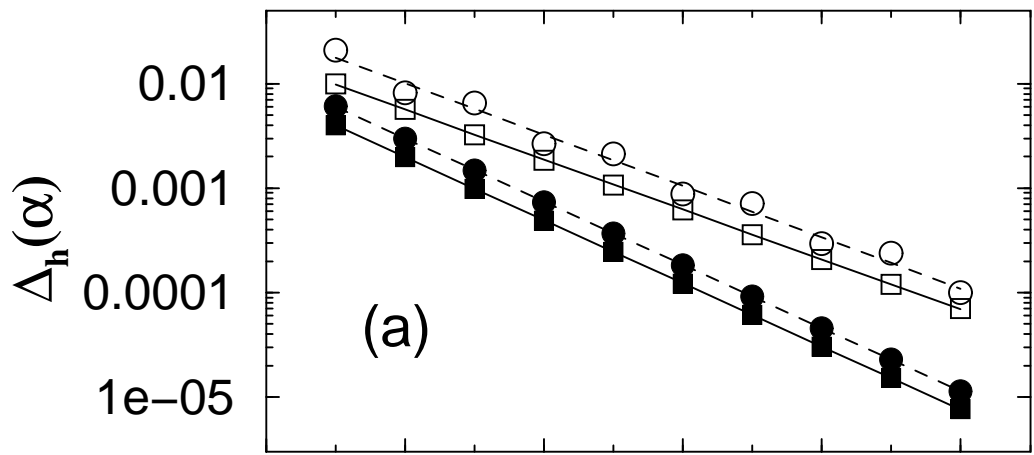
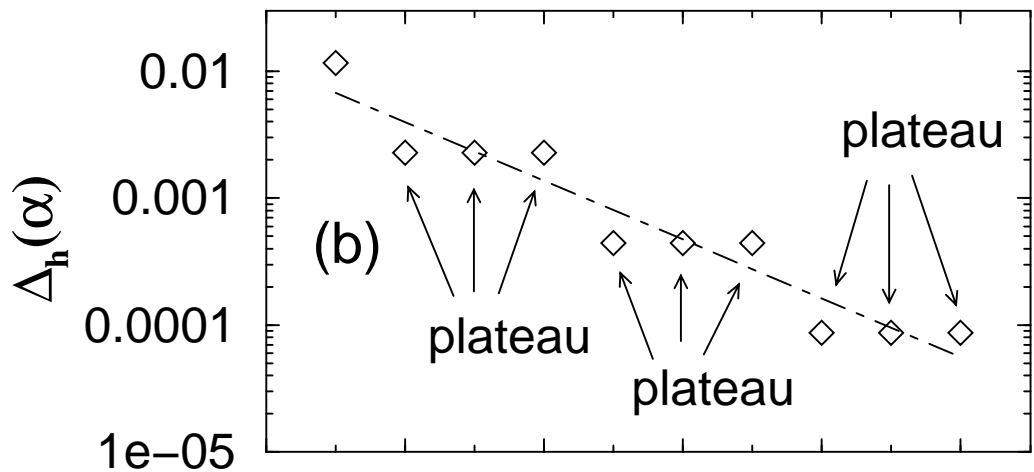
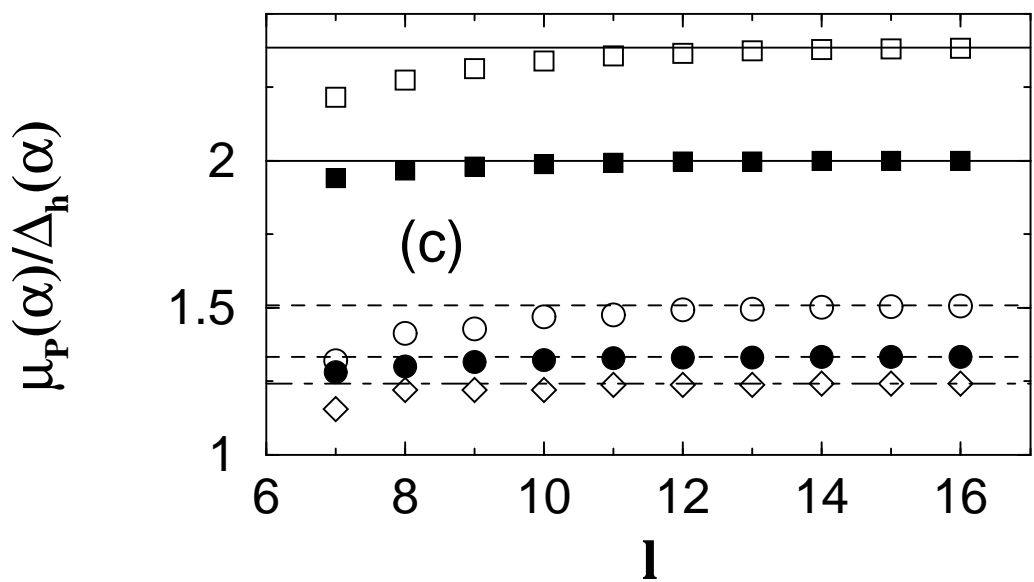
\end{document}